\documentclass{paper}
\usepackage{amsmath,epsf}
\usepackage[T1]{fontenc}
\usepackage[latin1]{inputenc}
\usepackage{epsf}
\usepackage{amsfonts,amssymb}
\usepackage[usenames,dvipsnames]{color}
\begin{document}

\begin{center}
{\Large Renormalization Theory based on Flow Equations}
\end{center}
\begin{center}
{\it Lecture given at the 
symposion in honour of Jacques Bros, 19-21 July 2004, Saclay} 
\end{center}

\begin{center}
{\large \baselineskip 20pt
Christoph Kopper\footnote{email~: 
kopper@cpht.polytechnique.fr} 
}
\end{center}

\centerline{Centre de Physique Th{\'e}orique de l'Ecole Polytechnique}
\centerline{F-91128 Palaiseau, France}
\date{october 2004 }

\begin{abstract}
\normalsize
I give an overview over some work on rigorous renormalization
theory based on the differential flow equations of the  Wilson-Wegner 
renormalization group. I first consider
 massive Euclidean $\varphi_4^4$-theory.
The renormalization proofs are achieved through
inductive bounds on regularized Schwinger functions.
I present relatively crude bounds which are easily proven,
and sharpened versions (which seem to be optimal as regards 
large momentum behaviour). Then renormalizability statements in Minkowski 
space are presented together with analyticity properties
of the Schwinger functions.
Finally I give a short description of further results.

\end{abstract}

\noindent
PACS classification 11.10.Gh

\newpage

\section{Introduction }
\vskip.4cm\noindent
In this lecture I would like to give a short 
overview over part of the work on renormalization theory 
based on flow equations I have been involved in since 1990.
The differential flow equations of the Wilson renormalization  
group [1] appear for the first time in a paper of Wegner and Houghton  [2]
in 1972. In a seminal paper by Polchinski [3] in 1984 a renormalization
proof for scalar $\varphi_4^4$-theory was performed. It is  based on the 
observation that the flow equations give access to a tight inductive
scheme wherefrom bounds on the regularized Schwinger functions implying
renormalizability may be deduced. The Schwinger functions are
regularized by an $UV$-cutoff $\Lambda_0$ and by an infrared
cutoff $\Lambda$, the flow parameter. The bounds on these Schwinger 
functions obtained in [4] are {\it uniform in the UV cutoff} and finite
for $\Lambda \to 0$. This basically solves the renormalization
problem. 
Later on renormalization theory with flow equations was extended in
various directions, for a recent review covering some of the rigorous
work see [5].   
Among the issues treated are
the renormalization of composite operators
and the short distance expansion, sharp bounds on the Schwinger
functions, massless theories, 
the transition to Minkowski space, the treatment of abelian
and nonabelian gauge theories and field theories at finite temperatures. 

I will adress subsequently massive Euclidean $\varphi_4^4$-theory.
I first outline the simplest version of the renormalization proof
based on relatively crude bounds on the regularized Schwinger functions.
Then I present sharpened versions of these bounds in momentum and
position space. Hereafter   analyticity properties of the
Schwinger functions  and statements on renormalizability
in Minkowski space are presented.
These notes end with a short overview on further rigorous results
obtained in the flow equation framework.  
 
Colleagues I had the pleasure to work with on 
renormalization theory with flow equations are 
Fr{\'e}d{\'e}ric Meunier,
Walter Pedra,
Thomas Reisz, Manfred Salmhofer, Clemens Schophaus and Vladimir
Smirnov. In particular I am grateful for the fruitful long term 
collaborations with 
Georg Keller and Volkhard M{\"u}ller.
\section{RENORMALIZATION of $\varphi_4^4$-THEORY}
\vskip.4cm\noindent
Massive Euclidean $\varphi_4^4$-theory is the simplest theory on which
the general issues of renormalization theory can be studied. 
When one analyzes the divergence structure of its
bare Feynman amplitudes one faces the problem
to be solved by general renormalization theory, namely
the so-called overlapping divergences which had paved the way to a 
rigorous theory of renormalization with so many difficulties during the
first few decades of existence of perturbative quantum field theory.    
The most astonishing message from the flow equation framework is that
the focus on overlapping divergences and the heavy combinatorial
and analytical machinery employed for its solution 
are not intrinsic to the problem\footnote{This is of course is of no
consequence as for the merit attached to the  hard and profound work
done in early rigorous renormalization theory.}. In the flow equation
approach  overlapping divergences  do not leave any trace.

\subsection{The basic tools}

The bare propagator of the theory is replaced 
by a regularized {\it flowing} propagator
\[
C^{\Lambda,\Lambda_0}(p)\,=\, {1 \over  p^2+m^2} \
\{ e^{- {p^2+m^2 \over \Lambda_0^2}} 
-e^{- {p^2+m^2 \over \Lambda^2}} \}\ , \quad
0 \le \Lambda \le \Lambda_0 \le \infty \ .
\]
The full propagator is recovered by taking the regulator 
$\Lambda_0 $ to $\infty$ and the flow parameter $\Lambda$
to $0\,$.
We calculate the derivative of $C^{\Lambda,\Lambda_0 }\,$ as
\[
\dot{C}^{\Lambda}(p)\,=\,\partial_{\Lambda}C^{\Lambda,\Lambda_0 }(p)\,=\, 
 -\,{2 \over  \Lambda ^3} \, e^{- {p^2+m^2 \over \Lambda^2}} \ .
\] 
Subsequently we will denote by
$\,d\mu_{\Lambda,\Lambda_0}$ the 
Gaussian measure with covariance $\hbar \,C^{\Lambda,\Lambda_0 }$.
The parameter $\hbar$ is introduced as usual to obtain a systematic
expansion in the number of loops.\\
The theory we want to study is massive Euclidean $\varphi_4^4$-theory. 
This means that we start from the {\it bare action}
\[
  L_0(\phi) = \int \! \! d^4 x \, \{ {g \over 4!} \, \varphi^4  
\,+ \, a_0 \,\varphi^2 +
     b_0\,  (\partial_{\mu}\, \varphi)^2 +
    c_0 \,\varphi^4\}
\]  
\[
a_0\,,\  c_0 =O(\hbar)\,,\quad b_0 =O(\hbar^2)\ .
\]
From the bare action and the flowing propagator we may define 
Wilson's {\it flowing effective action} $L^{\Lambda,\Lambda_0 }$ 
by integrating out momenta in the region 
$\Lambda ^2 \le p^2 \le \Lambda_0  ^2$.
It is defined through
\[
e^{- {1 \over \hbar}L^{\Lambda,\Lambda_0}(\varphi)}
~:=\, {\cal N}\
\int \, d\mu_{\Lambda,\Lambda_0}(\phi) \; 
e^{- {1 \over \hbar}L_0(\phi\,+\,\varphi)}
\]
and can be recognized to be the generating functional of the connected
free propagator amputated Schwinger functions of the theory with
propagator $C^{\Lambda,\Lambda_0 }$ and bare action $L_0\,$. 
For the normalization factor $  {\cal N}$
to be finite we have to restrict the theory to finite volume. All
subsequent formulae are valid also in the thermodynamic limit since
they do not involve any more the vacuum functional or partition function.

The fundamental tool for our study of the renormalization problem 
is then the functional  {\it Flow Equation}
\[
\partial_{\Lambda}\,L^{\Lambda,\Lambda_0}\,=\, 
\frac{\hbar}{2}\,
\langle\frac{\delta}{\delta \varphi},\dot{C}^{\Lambda }\,
\frac{\delta}{\delta \varphi}\rangle L^{\Lambda,\Lambda_0}
\,-\,
\frac{1}{2}\, \langle \frac{\delta L^{\Lambda,\Lambda_0} }{\delta
  \varphi} ,\dot{C}^{\Lambda } \,
\frac{\delta L^{\Lambda,\Lambda_0}}{\delta \varphi}\rangle\ .
\]
It is obtained by deriving both sides of the 
previous equation w.r.t. $\Lambda$ and performing an integration by parts
in the functional integral on the r.h.s. 
We then expand $L^{\Lambda,\Lambda_0}$ in moments w.r.t. $\varphi$
\[
(2\pi)^{4(n-1)}\,\,\delta_{\varphi(p_1)} \ldots \delta_{\varphi(p_n)}
L^{\Lambda,\Lambda_0}|_{\varphi\equiv 0}
\ =
\ \delta^{(4)} (p_1+\ldots+p_{n})\, 
{\cal L}^{\Lambda,\Lambda_0}_{n}(p_1,\ldots,p_{n})
\]
and also in a formal powers series w.r.t. $\hbar\,$ to select the loop
order $l$
\[
{\cal L}^{\Lambda,\Lambda_0}_{n}\,=\,
\sum_{l= 0}^{\infty} \hbar^l\,{\cal L}^{\Lambda,\Lambda_0}_{l,n}\ .
\]
From the functional flow equation we then obtain the perturbative
flow equations for the (connected free propagator amputated) 
n-point functions by identifying
coefficients 
\[
\partial_{\Lambda} \partial^w \,{\cal L}^{\Lambda,\Lambda_0}_{l,n} =
{1 \over 2}  \int_k \partial ^w 
{\cal L}^{\Lambda,\Lambda_0}_{l-1,n+2}(k,-k,\ldots)\ \dot{C}^{\Lambda} 
\]
\[
-
\sum_{l_i, n_i, w_i} c_{\{w_i\}} 
\Biggl[ \partial^{w_1} {\cal L}^{\Lambda,\Lambda_0}_{l_1,n_1}
\,\,(\partial^{w_3}\dot{C}^{\Lambda})\,\,
\partial^{w_2} {\cal L}^{\Lambda,\Lambda_0}_{l_2,n_2}\Biggr]_{sym}\ ,
\]
\[
l_1+l_2 =l\,, \quad
n_1 + n_2 =n+2 \,,\quad
w_1 + w_2 +w_3 =w\ .
\]

\begin{figure}
\centerline{\mbox{\epsfysize 2cm \epsffile{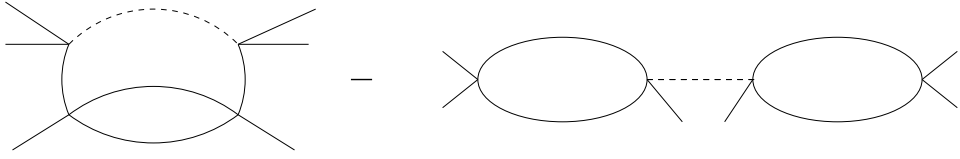}}}
\caption{A contribution to the r.h.s. of the flow equation for
  $l=2,\,n=6\,$.
The dashed line represents the derived propagator $\dot{C}^{\Lambda}\,$.}
\end{figure}

Here we wrote the equation directly in a form where 
a number $|w|$ of momentum derivatives, characterized by a multi-index $w$,
 act on both sides.
Derived Schwinger functions are needed to make close the inductive
scheme (see below). The $c_{\{w_i\}}$ are combinatoric constants
The subscript $sym$ indicates a symetrization procedure w.r.t.
external momenta.
\newpage
\subsection{Renormalizability}
Before bounding the solutions of the system of flow equations we first 
have to specify the boundary conditions~:\\
At $\Lambda = \Lambda_0 $ we find as a consequence of our choice of the
bare action $L_0 =\ L^{\Lambda_0,\Lambda_0} $
\[ 
\partial^w  {\cal L}^{\Lambda_0,\Lambda_0}_{l,n}\equiv 0\ \ \mbox{ for }\
n+|w|\ge 5\ .
\]
The so-called relevant parameters of the theory or renormalization
constants are explicitly fixed by renormalization
conditions imposed for the fully integrated theory 
at $\Lambda = 0\,$:\\[.2cm]
\[
 {\cal L}^{0,\Lambda_0} _4(0)=\,g \,,\quad
 {\cal L}^{0,\Lambda_0}_2(0)=\,0 \,,\quad
 \partial_{p^2}  {\cal L}^{0,\Lambda_0}_2(0)=\,0  
\]
(where we chose for simlicity  BPHZ renormalization conditions).\\
Once the boundary conditions are specified the renormalization
problem can be solved {\it inductively} as follows~:\\[.1cm]
The {\it inductive scheme} may for example be chosen as in the
subequent figure, namely by ascending  in $n+2l$ 
and for fixed $n+2l$ ascending in $l$. For this scheme to work
it is important to note that by definition there is no $0$-loop 
two-point function in $L^{\Lambda,\Lambda_0}\,$.

\begin{figure}[h]
\centerline{\mbox{\epsfysize 5.5cm \epsffile{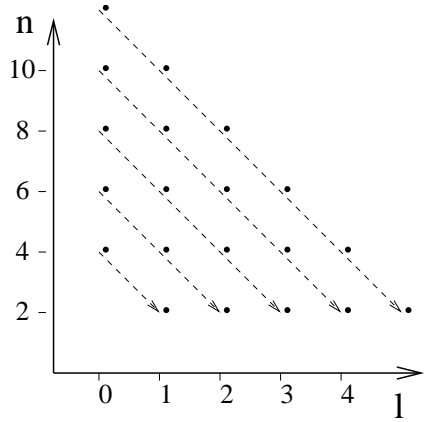}}}
\caption{The inductive scheme
starts from the point (l=0,\,n=4) and follows the arrows,
along the lines of fixed value of $n+2l\,$ in increasing order.}
\end{figure}

\noindent
We then may verify the follwing\\[.1cm]
{\it Induction hypothesis~:}

\[
|\partial^w {\cal L}^{\Lambda,\Lambda_0}_{l,n}(\vec{p})| \le
(\Lambda_m)^{4-n-|w| }\,{{\cal P}_1}(\ln {\Lambda_m  \over m})\,
{{\cal P}_2}({|\vec{p}| \over \Lambda_m})\ .
\]
Here we note
$\ \vec{p}=(p_1,\ldots,p_{n})\,,$\
$\ |\vec{p}|=\sup \{|p_1|,\ldots,|p_n|\}\,,$
${\ \,\,} \Lambda_m \,=\, \sup(\Lambda,m)\,$,
and the $\ {\cal P}_i$ are (each time they appear possibly new)
suitable polynomials 
with {\it coefficients}, which 
 {\it  do not depend on $\Lambda_0 $}. 
The important point to note on this induction hypothesis is that
it is {\it independent of (and thus uniform in) the UV cutoff} $\Lambda_0 \,$.
This means that its verification implies 
essentially the solution of the
renormalization problem. We shortly describe the\\[.1cm]
{\it Method of proof~:}\\
The induction hypothesis is proven by first verifying it for the boundary
conditions specified previously and then by inserting it on the right
hand side of the flow equation bounding the integrals in an elementary
way. On has to distinguish {\it irrelevant terms  with $n +|w| \ge 5$ 
which are integrated downwards from $\Lambda_0 $ to $\Lambda\,$
and relevant terms which are integrated upwards from $0$ to
$\,\Lambda\,$}. We note that the relevant terms are imposed by 
the renormalization
conditions at {\it fixed} external momentum. To move away from the
renormalization point one uses the Schl{\"o}milch interpolation
formula, e.g. for the two-point function

\noindent
\[
  {\cal L}^{\Lambda,\Lambda_0}_{l,2}(p)=\ 
 {\cal L}^{\Lambda,\Lambda_0}_{l,2}(0)\,+\,
\sum_{\mu}p_{\mu} \int_0^1 d\lambda \   (\frac{\partial}{\partial p_{\mu}}
 {\cal L}^{\Lambda,\Lambda_0}_{l,2})\,(\lambda p)\ , 
\]
and similar formulas where ${\cal L}^{\Lambda,\Lambda_0}_{l,2}(p)\,$ 
is replaced by
$\frac{\partial}{\partial p_{\mu}}
 {\cal L}^{\Lambda,\Lambda_0}_{l,2}(p)\,$ or by $\frac{\partial ^2}{\partial 
p_{\mu}\partial p_{\nu}}
 {\cal L}^{\Lambda,\Lambda_0}_{l,2}(p)\,$. When applying three
derivatives to the two-point function or one derivative to the 
four-point function the contribution becomes irrelevant and is
integrated downwards from $\Lambda_0 \,$. 
In this way the analogue of the previous equation 
written for $\frac{\partial ^2}
{\partial p_{\mu}\partial p_{\nu}}
 {\cal L}^{\Lambda,\Lambda_0}_{l,2}(p)\,$  gives full control of the 
twice derived two-point function. Then the equation for 
$\frac{\partial}{\partial p_{\mu}}
 {\cal L}^{\Lambda,\Lambda_0}_{l,2}(p)\,$ gives control of 
the once derived two-point
 function, and finally the one for  ${\cal L}^{\Lambda,\Lambda_0}_{l,2}(p)\,$ 
gives control of the two-point
 function itself. By Euclidean invariance one realizes that no 
renormalization conditions are needed for terms which are not scalars w.r.t.
this symmetry.
For the  four-point function only one
step is required.
\\
As an example we now bound by induction the first term on the r.h.s.
of the flow equation in the irrelevant case, i.e. for $n+|w| \ge 5$\\[.3cm]
\[
\int_{\Lambda}^{\Lambda_0} d\lambda \int_k\,
\frac{2}{\lambda ^3} \ e ^{-\frac{k^2+m^2}{\lambda ^2}}\ 
\lambda_m^{4-(n+2)-|w|}
\,\,{{\cal P}_1}(\ln {\lambda_m  \over m})\,
{{\cal P}_2}({|(\vec{p},k,-k)| \over \lambda_m})\
\]
\[
\le 
\int_{\Lambda}^{\Lambda_0} d\lambda\ \lambda_m^{4+1-(n+2)-|w|}\,
\,{{\cal P}_1}(\ln {\lambda_m  \over m})\,
{\tilde{\cal P}_2}({|\vec{p}| \over \lambda_m})
\ \le \
(\Lambda_m)^{4-n-|w|}\
{\tilde{\cal P}}_1(\ln {\Lambda_m  \over m})\
{\tilde{\cal P}_2}({|\vec{p}| \over \Lambda_m})
\]
which verifies the induction hypothesis.

\noindent
To complete the proof of renormalizability one also  proves 
\[
|\partial_{\Lambda_0}\partial^w {\cal L}^{\Lambda,\Lambda_0}_{l,n}
(\vec{p})| \le\,
{1\over \Lambda_0^2}(\Lambda_m)^{5-|n|-|w| }\,{{\cal P}_1}(\ln (\Lambda_0/m))\,
{{\cal P}_2}({|\vec{p}| \over \Lambda_m})\ .
\]
This bound is obtained applying the same inductive scheme on the flow 
equation derived once w.r.t. $\Lambda_0\,$ and using the previous
bounds on $\partial^w {\cal L}^{\Lambda,\Lambda_0}_{l,n}
(\vec{p})$. It implies convergence of the 
$\partial^w {\cal L}^{\Lambda,\Lambda_0}_{l,n}
(\vec{p})$
for $\Lambda_0 \to \infty\,$.
whereas the previous  induction hypothesis still allows for
solutions of the flow equations which are uniformly bounded but 
oscillating in terms of $\Lambda_0 \,$.

\subsection{Generalizations and Improvements}

The previous statement of renormalizability may be improved and
generalized in various ways. A first short remark is that we may 
generalize the boundary conditions by enlarging the class
of bare actions~: we also admit {\it nonvanishing irrelevant 
terms in the bare action} as long as they satisfy the bound
\[
| \partial^w  {\cal L}^{\Lambda_0,\Lambda_0}_{l,n}(\vec{p})|\,\le
\Lambda_0^{4-n-|w| }\,{{\cal P}_1}(\ln {\Lambda_0  \over m})\,
{{\cal P}_2}({|\vec{p}| \over \Lambda_0})\
\mbox{ for }
n+|w|\ge 5\ .
\]
Remembering the previous method of proof we only have to note that
these boundary conditions still satisfy the induction hypothesis
(for $\Lambda =\ \Lambda_0$) so that the proof goes through without change.
One might note that this generalization is of practical importance
on one hand, since such irrelevant terms typically will appear, e.g.
in effective actions used to describe critical behaviour in statistical
mechanics [6]. On can calculate subdominant corrections
to scaling near second order phase transitions
due to such terms.\\
On the other hand these terms pose considerable problems in the 
BPHZ type renormalization proofs of which we are not sure whether they
have been solved in full rigour up the present day [7].\\   

It is also possible to give much {\it sharper bounds} on the Schwinger
functions with the aid of the flow equations which do not only imply
renormalizability but also restrict the high momentum or large
distance behaviour of 
the Schwinger functions  in a basically optimal way (in the sense that
they are saturated by individual Feynman amplutudes). To phrase those 
bounds we need the following concept of\\[.2cm] 
{\it Weighted trees}:\\
We regard trees with $n\,$ external
lines and $\cal V$ vertices of coordination number 3 or 4.
To the $n\,$ external lines of a tree $T$ 
we associate $n\,$ external incoming momenta
$\vec p=\,(p_1,\ldots,p_n\,)\,$ respectively positions
$\vec x=\,(x_1,\ldots,x_n\,)\,$.
We then define weight factors in momentum and position
space.\\[.2cm]
{\it Weight factor in momentum space~:}\\
To each internal line $I \in {\cal I}\, $ of the tree is attached
a weight  $\mu(I)\in \{1,2\}$ 
such that $\,\sum_{I\in{\cal I}}\mu(I)=n-4\,$.
Let $p(I)\,$ be the 
momentum flowing through the internal line $I\in \cal I\,$.
For given $\Lambda\,$ and  tree $T\,$ we define
\[
g^{\Lambda}( T)=\,
\prod_{I\in{\cal I}}{1 \over (\sup(\Lambda_m,|p(I)|)^{\mu(I)}}\ .
\]
\\[.1cm]
{\it Weight factor in position space~:}\\
For $\alpha=\, 1-\varepsilon\,$ fixed  and letting $|I|\,$ be the distance
of the points joined by $I\,$ in position space we define
\[
{\cal F}_{\Lambda}(T)
~:= \Lambda_m^{4-n}\, \prod_{I\in {\cal I}}\
e ^{-[\Lambda_m \,|I|]^{\alpha}}\ .
\]
In terms of these weight factors we may now state\\[.1cm] 
{\it Sharp bounds in momentum space~:} [8]
\[
|{\cal L}_{l,n}^{\Lambda,\Lambda_0}({\vec p})|
\leq
\sup_T \,g^{\Lambda}(T)\
{\cal P}_{l}\bigl(\ln\sup({|\vec p| \over \Lambda_m},
{\Lambda_m \over m})\bigr)\,,
\ n \geq 4\,,
\]
\[
  |{\cal L}_{l,2}^{\Lambda,\Lambda_0}(p)|\leq
            \sup(|p|,\Lambda_m)^{2}\
{\cal P}_{l-1}\bigl(\ln\sup({|p|\over \Lambda_m},{\Lambda_m\over
       m})\bigr) \ ,
\]
\[
 \deg{\cal P}_{l}\le\,\,l\ .
\]
The most important ingredient to obtain such sharpened bounds is to
implement in the inductive procedure the fact that momentum
derivatives improve the large momentum behaviour.
These derivatives necessarily appear due to the use 
of the Schl\"omilch interpolation formula. 
Here the main problem to solve is to find optimal paths in the
interpolation formula for the four point function so as to avoid
that derivatives only lead to a net gain of a small external momentum only
whereas subsequent application of the interpolation formula  
then may lead to multiplication by a
large one. Thus the problem is related to the exceptional momentum
problem.\\[.2cm]  
{\it Sharp bounds in position space~:}\\
On smearing  out the positions of $n-1$ external vertices 
with smooth bounded test functions  $\varphi~: =\,
\varphi_2\cdot \ldots \cdot \varphi_n\,$,
$\varphi_i$ being supported in a square of side length $1/\Lambda_m$
around $x_i$,
we have for suitable (maximizing) choice of the  positions of the internal
vertices of the trees for $n \ge 4$
\[
| \, {\cal L}^{\Lambda,\Lambda_0}_{l,n} (\varphi,x_1)|\,\le\ \sup_{T} 
 {\cal F}_{\Lambda}(T)\
{\cal P}_l(\ln (\varphi,\Lambda_m ))\,  ||\varphi||\ ,
\]
\[
| \,  {\cal L}^{\Lambda,\Lambda_0}_{l,2} (\varphi,x_1)|\,\le\, 
\Bigl(||\varphi||\,+ \,
 \Lambda_m^{-2} \,  ||\varphi^{(2)}||\,+\,
 \Lambda_m ^{-3}\, ||\varphi^{(3)}||\Bigr) 
\ e ^{-[\Lambda_m \,dist(x_1,\, supp \varphi)]^{\alpha}}
\ {\cal P}_{l-1}(\ln (\varphi,\Lambda_m )) \ .
\]
Here the definiton of ${\cal P}_{l-1}(\ln (\varphi,\Lambda_m )) $ 
is given through
\[
{\cal P}_{l}(\ln (\varphi,\Lambda_m) ):=\
{\cal P}_{l}(\ln \sup( {||\varphi^{'}|| \over ||\varphi|| \Lambda_m },
\,{\Lambda_m \over m}))\ ,
\]
with the definition $||\varphi ^{(k)}||:=\ \sup_{w,|w| \le
  k}||\varphi ^{(w)}||_{\infty}\,$. 
This means that the polynomial is in logarithms 
of the maximal momentum content of the test functions if the latter
exceeds $\Lambda_m/m\,$.\\
The proof of these bounds has not been published so far.\\[.2cm]  
{\it Sharp combinatoric bounds~:} [9]\\
Finally the flow equation method also permits to recover
the bounds on large orders of perturbation theory  
which were established by de Calan and Rivasseau using
BPHZ type methods [10]. 
They can be stated now in a  form somewhat more compact
than the one given in the original paper [9]:\\
\[
|{\cal L}^{\Lambda,\Lambda_0}_{l,n}(\vec{p})| \, \le \, \Lambda_m^{4-n}\ 
K^{{n-3\over 2}+2l} \
 ({n \over 2})!\ 
\sum_{s=0}^l ({n \over 2} + s-2)!\ \ln^{l-s}({\Lambda_m \over m}) 
\]
\[
\mbox{ for } \ |\vec p| \le \sup(2\Lambda_m, k)\, \mbox{ with }\ k\,
\mbox{ fixed, and for } K\,  \mbox{ sufficiently large.}
\]

\section{Relativistic Theory}

We want to indicate how the flow equation method, 
which at first sight is restricted to euclidean 
theories, can be employed to prove renormalizability in 
Minkowski space [11]. In this respect we emphasize the beautiful framework
for renormalization theory developed by Epstein and Glaser, 
which takes into account the structural locality and causality 
properties of relativistic theories from the beginning, and which
was presented at the Saclay conference 
in the contributions of Dorothea Bahns
and Klaus Fredenhagen.  
In our framework we address the relativistic renormalization problem
in momentum space, and we want to recover the results from BPHZ
renormalization theory saying that the perturbative relativistic 
Green functions are distributions in momentum space, and at the same
time analytic functions in those domains where the external incoming
energy stays below the physical thresholds. For a fully satisfactory 
renormalization theory in Minkowski space one would also like to have
continuity properties of the Green functions above threshold,
sufficient at least to define a physical renormalized coupling
for physical values of the external momenta which necessarily lie 
above threshold. Satisfactory results 
in this respect do not seem to exist, and we hope to come back on 
this issue in the future.

\subsection{ The Flow Equation for   one particle irreducible
Schwinger functions}

 To discuss analyticity
properties it is preferable to work with one particle irreducible
(1PI) Schwinger functions, the generating functional of which is obtained from
the one for connected Schwinger functions by a Legendre transform.
Using this relation one can deduce the (still Euclidean) flow equations 
for the perturbative 1PI Schwinger functions 
$\Gamma^{\Lambda,\Lambda_0}_{l,n}\,$
which take the following form  
\[
\partial_{\Lambda} \,\Gamma^{\Lambda,\Lambda_0}_{l,n} =
{1 \over 2}  \int_k  
{\hat \Gamma}^{\Lambda,\Lambda_0}_{l-1,n+2}(k,-k,\ldots)\ 
\dot{C}^{\Lambda,\Lambda_0} (k)
\]
\[
{\hat \Gamma}^{\Lambda,\Lambda_0}_{l,n} =
\sum_{i \ge 1}(-1)^{i+1} \sum_{l_i, n_i}\ \Bigl[ \Bigl(\,\prod_{j=1}^{i-1}  
\ {C}^{\Lambda,\Lambda_0}(k_j) \ 
\Gamma^{\Lambda,\Lambda_0}_{l_j,n_j+2}\, \Bigr)\
\Gamma^{\Lambda,\Lambda_0}_{l_i,n_i+2}\,\Bigr]_{sym} \ \,,
\]
\[
\sum_i l_i = l\ ,\quad \sum_i n_i =n-2\, (1-\delta_{i,1})\ .
\]
The momentum arguments $k_j$ are determined by momentum conservation.
The are a sum of the loop momentum $k$ and a subsum of incoming
momenta $p_i$
\begin{figure}
\centerline{\mbox{\epsfysize 3.6cm \epsffile{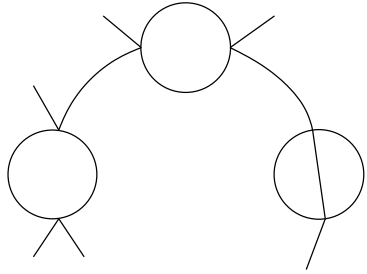}}}
\caption{A contribution to $\ {\hat \Gamma}_{l,n}$ for 
$l=4,\ n=6\,$.}
\end{figure}
One immediately realizes that the inductive scheme used for the
renormalization proof of connected Schwinger functions 
is also viable in the 1PI case, and the same inductive bounds
may also be proven for  the euclidean 1PI functions
$\Gamma^{\Lambda}_{l,n}(\vec p)\,$. 
\subsection{Analyticity of the Euclidean theory}
Using the 1PI flow equation it possible to extend 
the inductive renormalizability proof to the  
{\it  complex domain} 
\[
{\cal D} = \Bigl\{(p_{01},\underline{p}_1,
\ldots,p_{0n-1},\underline{p}_{n-1}) \,|\
\underline{p}_i \in {\mathbf R}^3\ ,\
p_{0i} \in {\mathbf C} \ , \quad
|Im\sum_{i\in J } p_{0i} | < 2m- \eta \quad  \forall \,J \subset \{1,\ldots,
 n\}\, \Bigr\}
\]
\centerline{(with $\eta>0\,$ arbitrarily small)}\\
that is to say~:\\
 {\it The 1PI Schwinger functions  $\Gamma^{\Lambda,\Lambda_0}_{l,n}(\vec
p)\,$ are analytic  in ${\cal D} \,$
with respect to the arguments $\,p_{01},\ldots, p_{0n}\,$}
and still uniformly bounded with respect to $\Lambda_0$
by bounds analogous to those given for real external momenta
(for $\eta>0\,$ fixed).\\ 
The proof of this statement obviously uses the fact 
that we have chosen an analytic regulator
in momentum space. It relies on the observation that the external
momentum sets  of the terms $\Gamma^{\Lambda,\Lambda_0}_{l_i,n_i+2}\,$ 
appearing on the r.h.s. of the 1PI flow equation
also fulfill
the conditions appearing in the definition of $\cal D\,$, if the 
full external momentum set does. Thus the inductive procedure 
is still valid. Then by displacing the integration  contour 
of the first component of the
integration variable $k\,$ in the flow equation by an amount of
modulus smaller than $m\,$ in the imaginary direction 
- in fact one has to move towards  the centre of the interval of width 
$2m -\eta\,$ appearing in the definition of $\cal D\,$ - 
one can show that all propagators ${C}^{\Lambda,\Lambda_0}(k_i)\,$ 
in the 1PI flow equation have  still $|Im\, k_{0i}| < m-\eta\,$ 
so that the $k\,$-integral is still controlled by the exponential
fall-off \, $\exp(-\frac{k^2_i+m^2}{\Lambda ^2})\,$ of the regulating factors. 
We note again that the bounds thus
obtained are (obviously) not uniform in $\eta\,$.

To obtain the statement on the existence of the relativistic Green
functions as distributions we need more explicit information
on the dependence of the 1PI Schwinger functions on the external momenta.  
This information is obtained from the following 
$\alpha$-parametric integral representation 
for these Schwinger functions
\[
\Gamma^{\Lambda,\Lambda_0}_{l,n}(\vec p) =\
\int_{1/ \Lambda_0 ^2}^{1/\Lambda ^2} d \alpha_1 \ldots d \alpha_s\
G_{l,n}^{\Lambda,\Lambda_0}(\vec \alpha , \vec p)
\]
with
\[
G^{\Lambda,\Lambda_0}_{l,n}(\vec{\alpha},\vec{p})\,=\,
\sum_j V_j(\vec{\alpha}) \,P_j(\vec{p})\,Q_j(\vec{\alpha})\,
e^{-(\vec{p},A_j\,\vec{p})_{eu}-m^2\sum_{k=1}^s\alpha_k }\ .
\]
By induction one verifies the following statements~:\\
(i) $P_j$ is a monomial  in the $O(4)$-invariant 
scalar products of the external momenta.\\  
(ii) $Q_j$ is a rational function homogeneous of degree
$d_j > -s\,$ in
the $\alpha_i$. Here $s= 2l -2+ n/2$ is the number of internal lines 
of $\Gamma_{l,n}$.\\
(iii) $A_j$ is  a positive-semidefinite symmetric 
($n\!-\!1\!\times\!n\!-\!1$)-matrix,
homogeneous of degree $1$ in $\vec{\alpha}$.\\
(iv) $V_j$ is a product of $\theta$-functions restricting
the $\alpha$-integration domain. Its origin traces back to previous
integrations over the flow parameter,
which either start from $\Lambda_0$ for the irrelevant terms or from
$0$ (correspon\-ding to $\alpha \to \infty$) for the relevant terms.\\
On changing variables
\[
\alpha_k\,=\,\tau \beta_k\,,\quad
d\vec{\alpha}\,=\,\tau^{s-1}\delta(1-\sum\beta_k)d\vec{\beta}\,d\tau
\]
and {\it using homogeneity of  $A_j$ und $Q_j$} we may write 
\[
\Gamma ^{\Lambda,\Lambda_0}_{l,n}(\vec{p})\,=\,
\int_0^1 d\vec{\beta}\,\delta(1-\sum\beta_k)
\sum_jV_j(\vec{\beta})\ P_j(\vec{p})\ Q_j(\vec{\beta})
\ \frac{1}{[(\vec{p},A_j\,\vec{p})_{eu}+m^2]^{d_j+s}}\ .
\]
The integrand is absolutely  integrable with respect to  $\vec \beta\,$
for $\vec p \,\in \,\cal D\,$. This statement follows from the
previous one on the degree of $Q_j$.\\
The previous  integral representation is obtained quite naturally from the 
starting observation that
\[
 {C}^{\Lambda,\Lambda_0}(k)=\ \int_{1/\Lambda_0 ^2}^{1/\Lambda ^2} 
e ^{-\alpha(k^2
   +m^2)}\ 
d\alpha
\]
and on using recursively (inductively) the flow equation
where each successive $\Lambda\,$-integration is written as a
new $\alpha$-integration. 

\subsection{Transition to Minkowski space}

We use the relativistic  Feynman propagator 
\[
C_{rel}(p)\,=\,\frac{i}{p^2_{rel}-m^2+i\varepsilon(p^2_{eu}+m^2)}
\]
\[
C^{\Lambda,\Lambda_0}_{rel}(p)\,=\, \int_{\Lambda_0^{-2}}^{\Lambda^{-2}}
e^{i\alpha[p^2_{rel}-m^2+i\varepsilon(p_{eu}^2+m^2)]}
\,d\alpha \ ,
\]
which for finite $\varepsilon$ can be bounded in terms of the Euclidean one
and $1/\varepsilon\,$ to make all integrals well-defined.
Then following the steps which led to the integral representation for
the Euclidean theory one obtains the 
same representation for the relativistic theory, replacing
Euclidean by Minkowski scalar products and
\[
\frac{1}{(\vec{p},A_j\,\vec{p})_{eu}+m^2}
\]
by
\[
\frac{1}{(\vec{p}, A_j\, \vec{p})_{rel}
-m^2+i\varepsilon((\vec{p},A_j\, \vec{p})_{eu}+m^2)}\ .
\]
This representation together with results from distribution theory [12]
implies the following\\[.1cm] 
{\it Results}~:\\
1) The relativistic 1PI Green functions are  
 {\it  Lorentz-invariant tempered distributions}.\\
2) For external momenta $\,(p_{01},\underline{p}_1,\ldots,p_{0n},
\underline{p}_{n}) \,$
with  {\color{red} $|\sum_{i\in J } p_{0i} | < 2m$}
 $\, \forall J \subset \{1,\ldots,
 n\}$ they agree with the Euclidean ones
for $\,(ip_{01},\underline{p}_1,\ldots,ip_{0n},\underline{p}_{n}) \,$ 
and are thus
 {\it  smooth functions} 
in the (image of the) corresponding domain under the Lorentz
group.

\section{A short look on further results}

Among the results of rigorous renormalization theory with flow
equations not covered here we mention the following~:\\[.2cm]
1. One of the earliest applications of Polchinski's method
was by Mitter and Ramadas [13], who gave a renormalization proof for
the two-dimesnional nonlinear $\sigma$-model on the ultraviolet
side.\\[.2cm]  
2. Massless theories [14]\\
Massles theories or partially massless theories have been treated in
momentum space. In this case the inductive bounds 
of course have to allow for the singularities present in the connected
Schwinger functions at exceptional external momentum, i.e. when
subsums of external momenta vanish. One then has to find and prove  bounds 
which characterize these singularities in a basically optimal way so
that the successive integration procedures of the flow equation 
reveal how the possible singularities disappear when passing from
${\cal L}_{l-1,n+2}\, $ to ${\cal L}_{l,n}\,$. We think that a better
and more transparent way of treating this problem than that of [14]
is by the more recent method of [8]. There we give sharp large momentum
bounds for the massive theory. It should be possible to translate   
them to the massless theory, taking care of restrictions on the
admissible renormalization conditions in this case.\\[.2cm] 
2. Composite Operators, Zimmermann identities [15] and the 
Wilson short distance expansion [16]\\
The renormalization of Schwinger functions with composite operator 
insertions is an indispensable tool in many applications of renormalization
theory. In particular in gauge theories these composite operator
insertions appear as quantities which characterize the violation of
the Ward identities  due  to
the presence of cutoffs. Besides, inserted Green functions are of
physical interest in their own rigth, e.g. current-current-correlators  
in relativistic field theory, or 
the $<\varphi ^2\ \varphi ^2>$-correlator from which one reads
the behaviour of the specific heat when applying 
Euclidean field theory to critical phenomena.
The Zimmermann identities relate these composite operator insertions
in different renormalization schemes with each other. They are
obtained in a straightforward way from the linear structure of flow
equations for inserted Green functions. Having at hand the general composite
operator renormalziation techniques one can then prove
asymptotic expansions for perturbative Green functions of the Wilson
type. This has been done in Euclidean space only.\\[.2cm]   
3. Gauge theories [17, 18]\\
We mentioned that 
the Ward identities are violated
in momentum space regulated gauge theories. In the flow equation framework 
the renormalization of gauge theories then requires to prove that
the Ward identities can be restored in the limit where the cutoffs are 
taken away. To do so one disposes of the freedom of imposing
renormalization conditions on all relevant terms of the theory.   
This programme has been achieved explicitly for QED and for massive
SU(2) Yang-Mills-Higgs theory.\\[.2cm]  
4. Finite temperature theory [19]\\
Using flow equations it is possible to study explicitly
the difference between the (massive Euclidean scalar) finite temperature  
and zero temperature field theories and to show that this difference
theory is completely irrelevant in the sense of the renormalization
group. That is to say, both theories ($T>0$ and $T=0$) are
renormalzied with identical counter terms. This result is new
(though expected by experts) and its proof is 
simple~: our framework is particularly suited for this kind analysis
where one has to lay hand on the bare and renormalized actions at the
same time, since both appear automatically as opposite boundary values
of the flow to be controlled.  

Two further topics of related nature have been treated
within the flow equation framework.  First the
 Symanzik programme of improved improved actions [20]
has been carried out: it is possible 
to modify the bare action through irrelevant terms
such that the convergence of the regularized towards the renormalized
theory is  accelerated,  in terms of inverse powers 
of $1/\Lambda_0\,$. Secondly decoupling theorems [21] 
for heavy particles have been proven.\\[.5cm]   

\noindent
In conclusion we may say that\\[.2cm]
1. The  Wilson-Wegner  flow equation allows for a simple 
transparent rigorous
solution of the perturbative re\-normalization problem, without
introducing Feynman diagrams.\\
2. The method gives new results and perspectives 
on various aspects of the problem.\\
3. It places the problem in a more general, more physical
and less technical context,  
relating perturbative and nonperturbative
aspects.\\
We should also mention that
flow equations in different forms and in various approximation schemes 
are presently applied to many problems in high energy and  solid state   
physics, a topic not covered in this lecture. For a relatively recent
review see [22].
\vspace{1cm}

\noindent
Acknowledgement~: This manuscript was written when the author was a
visitor to DESY, Vienna. Support by ESI is gratefully acknowledged.

\end{document}